\documentclass{aa}
\usepackage{psfig}
\newcommand{\lsim}{\stackrel{\scriptstyle <}{\phantom{}_{\sim}}}
\newcommand{\gsim}{\stackrel{\scriptstyle >}{\phantom{}_{\sim}}}
\begin{document}

%   \thesaurus{08     % A&A Section 8: Stars
%              (02.04.1;  % Dense matter,
%               08.05.3;  %
%               08.09.3;  % Stars: interiors,
%               08.14.1;} % Stars: neutron.
%
   \title{Cooling of Neutron Stars and $3P_2$ Neutron Gap
        %\thanks{Research supported in part by the
%DFG under grant no. 436 RUS 17/117/03}
}

%   \subtitle{}

   \author{ H. Grigorian \inst{1,2}
\and D.N. Voskresensky \inst{3,4}}
%%}

 \offprints{D.N. Voskresensky}

 \institute{Institut f\"ur Physik, Universit\"at Rostock,
        Universit\"atsplatz 3, D--18051 Rostock, Germany\\
        email: hovik.grigorian@uni-rostock.de\\
         \and Department of Physics, Yerevan State University, Alex
        Manoogian Str. 1, 375025 Yerevan, Armenia\\
        \and Gesellschaft f\"ur Schwerionenforschung mbH, Planckstr. 1,
        D--64291 Darmstadt, Germany\\
        email: d.voskresensky@gsi.de\\
        \and Moscow Institute for Physics and Engineering, Kashirskoe sh. 31,
        115409 Moscow, Russia\\
%%email: d.voskresensky@gsi.de
\date{Received: 25. February 2004; accepted: }
}

\abstract{
We study the dependence of the cooling of isolated neutron
stars on the magnitude of the $3P_2$ neutron gap. It is demonstrated that
our ``nuclear medium cooling'' scenario is in favor of a suppressed value
of the $3P_2$ neutron gap.
}
\maketitle

%
%________________________________________________________________

\section{Introduction} \label{sec:intro}
As the result of many works the so called ``standard'' scenario of
neutron star (NS) cooling emerged, where the main process
responsible for the cooling  is the modified Urca process (MU)
$nn\rightarrow npe\bar{\nu}$ calculated using the free one pion
exchange between nucleons, see \cite{FM79}. This scenario explains
only the group of slow cooling data. To explain the group of rapid
cooling data the ``standard''  scenario was supplemented by one of the
so called ``exotic'' processes either with pion condensate, or
with kaon condensate, or with hyperons, or involving the direct
Urca (DU) reactions, see \cite{T79,ST83} and references therein. All
these processes may occur only for densities higher than a
critical density, $(2\div 6)~n_0$, depending on the model, where
$n_0$ is the nuclear saturation density.

Then the pair breaking and formation   (PBF) processes permitted
in nucleon superfluids have been suggested.  \cite{FRS76}
calculated the emissivity of the $1S_0$ neutron  pair breaking and
formation (nPBF) process  and \cite{VS87} considered a general
case. Neutron and proton  (pPBF) pair breaking and formation
processes  were incorporated within a closed diagram technique
including correlation effects. Up to factors of the order of 1
numerical estimates are valid both for $1S_0$ and $3P_2$
superfluids. \cite{SVSWW97} have shown that the inclusion of the
PBF processes into the cooling code may  allow to describe the
''intermediate cooling'' group of data (even if one artificially
suppressed medium effects). Thus the ''intermediate cooling''
scenario arose. Then the PBF  processes were incorporated  in the
cooling codes of other groups, elaborating the  ``standard plus
exotics'' scenario, see \cite{TTTTT02,YGKLP03,PLPS04}. Some papers
included the possibility of internal heating that results in a
slowing down of the cooling of old pulsars, see \cite{T04} and
Refs. therein. However, paying the price for a simplification of
the consideration, calculations being performed within the
``standard plus exotics'' scenario, did not incorporate in-medium
effects. Recently \cite{PLPS04} called this approach the ''minimal
cooling'' paradigm.

The necessity to include in-medium effects into the NS cooling
problem is a rather obvious issue. It is based on the whole
experience of condensed matter physics, of the physics of the
atomic nucleus and it is called for by the heavy ion collision
experiments, see \cite{MSTV90,RW,IKHV01}. The relevance of
in-medium effects for the NS cooling problem has been shown by
\cite{VS84}, (1986), (1987),
%%,VS86,VS87,
\cite{MSTV90,V00} and the efficiency of the developed  ``nuclear
medium cooling'' scenario for the description of the NS cooling
was demonstrated within the cooling code by \cite{SVSWW97} and
then by \cite{BGV04}. In that paper, it was shown that it is
possible to fit the whole set of cooling data available by today.
Besides the incorporation of in-medium effects into the pion
propagator and the vertices, it was also exploited that the $3P_2$
neutron gaps are dramatically suppressed. The latter assumption
 was motivated by the analysis of the data (see Figs 12, 15, 20 -- 23 of
 \cite{BGV04})
and by recent calculations of the $3P_2$ neutron gaps by \cite{SF03}.

%%%%%%%%%%%%%%%%%%%% Figure 5 %%%%%%%%%%%%%%%%%%
\begin{figure}[htb]
\psfig{figure=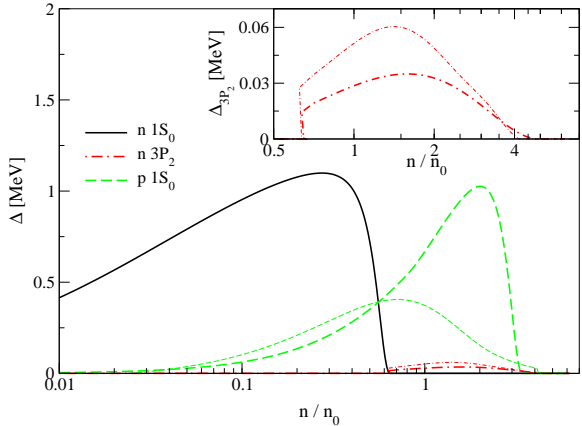,width=0.5\textwidth,angle=-90} \caption{
Neutron and proton pairing gaps according to model I
%%\cite{YGKLP03}
(thick solid,
dashed and dotted lines) and according to model II
%%\cite{TT04}
%%%{SVSWW97}
(thin lines), see text. The $1S_0$ neutron gap is the same in both
models, taken from  \cite{AWP}. \label{fig-gaps}}
\end{figure}
%%%%%%%%%%%%%%%%%%%%%%%%%%%%%%%%%%%%%%%%%%%%%%%%

%%\subsection{Nucleon superfluidity}

In spite of many calculations which have been performed, the
values of nucleon gaps in dense NS matter are poorly known. This
is the consequence of the exponential dependence of the gaps on
the density dependent potential of the in-medium $NN$ interaction.
This potential is not sufficiently well known. Gaps that we have
adopted in the framework of the ''nuclear medium cooling''
scenario, see \cite{BGV04}, are presented in Fig. \ref{fig-gaps}.
Thick dashed lines show proton gaps which were used in the work of
\cite{YGKLP03} performed in the framework of the ``standard plus
exotics'' scenario. In their model proton gaps are artificially
enhanced (that is not supported by  any microscopic calculations)
just to get a better fit of the data. We use their ``1p'' model.
Neutron $3P_2$ gaps presented in Fig. \ref{fig-gaps} (thick dash
-dotted lines) are the same, as those of ``3nt'' model of
\cite{YGKLP03}. We will call this choice, the model I. Thin lines
show $1S_0$ proton and $3P_2$ neutron gaps from \cite{TT04}, for
the model AV18 by \cite{WSS95} (we call it the model II). We take
the same $1S_0$ neutron gap  in both models I and II (thick solid
line), as it was calculated  by \cite{AWP} and was previously used
by \cite{SVSWW97} within the cooling code.
 \cite{BGV04} have used  models I and II
within the ``nuclear medium cooling'' scenario. As it was checked
there, since the  $1S_0$ neutron pairing gap exists only within
the crust, dying for baryon densities $n\geq 0.6~n_0$, its effect
on the cooling is rather minor. Opposite, the effect on the
cooling arising from the proton  $1S_0$ pairing and  from the
neutron $3P_2$ pairing, with gaps reaching up to rather high
densities, is pronounced. The NS cooling essentially depends on
the values of the gaps and on their density dependence. Findings
of \cite{SCLBL96,LS00}, who incorporated in-medium effects,
motivated us to check the possibility of rather suppressed $1S_0$
neutron and proton gaps. For that aim we  introduced pre-factors
for $1S_0$ neutron and proton gaps which we varied in the range
$0.2\div 1$, see Figs. 18 and 19 of \cite{BGV04}.

%%%%%%%%%%%%%%%%%%%% Figure 4 %%%%%%%%%%%%%%%%%%
\begin{figure}[htb]
\psfig{figure=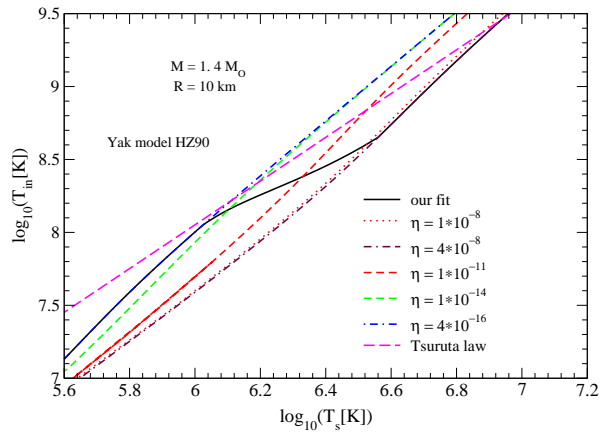,width=0.5\textwidth,angle=-90}
\caption{The relation between the inner crust temperature and the
surface temperature for different models. Dash-dotted curves
indicate boundaries of the uncertainty band. Notations of lines
are determined in the legend. For more details see  \cite{BGV04}
and \cite{YLPGC03}. \label{T-in} }
\end{figure}
%%%%%%%%%%%%%%%%%%%%%%%%%%%%%%%%%%%%%%%%%%%%%%%%

Recently \cite{SF03} have argued for a   strong suppression of the
$3P_2$ neutron gaps, down to values $\lsim 10~$keV, as the
consequence of the  medium-induced spin-orbit interaction. They
included important medium effects, as the modification of the
effective interaction of particles at the Fermi surface owing to
polarization contributions, with particular attention to
spin-dependent forces. In addition to the standard spin-spin,
tensor and spin-orbit forces, spin non-conserving effective
interactions were induced by screening in the particle-hole
channels. Furthermore a novel long-wavelength tensor force was
generated. The polarization contributions were computed to second
order in the low-momentum interaction $V_{\rm{low}\,k}$. These
findings motivated \cite{BGV04} to suppress values of $3P_2$ gaps
shown in Fig. \ref{fig-gaps} by an extra factor $f(3P_2 ,n)=0.1$.
Further possible suppression of the $3P_2$ gap is almost not
reflected on the behavior of the cooling curves.

%%%%%%%%%%%%%%%%%%%%%%%%%%% Figure 15  %%%%%%%%%%%%%%%%%%
\begin{figure}[htb]
\psfig{figure=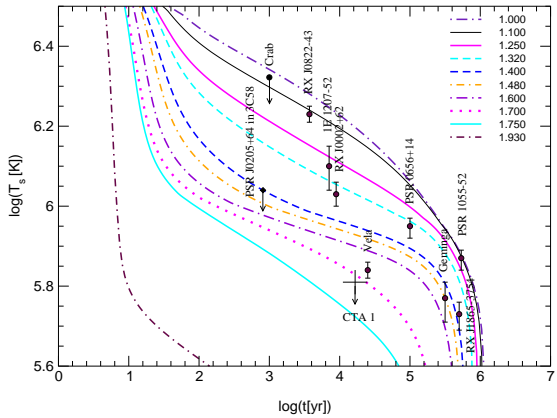,width=0.5\textwidth,angle=-90}
%%EV_HJSms.ps
\caption{Fig. 21 of \cite{BGV04}. Gaps are from Fig.  \ref{fig-gaps}
%%, thin line (Tamagaki and Takatsuka case).
for model II. The original $3P_2$ neutron
  pairing gap is  additionally suppressed by a factor $f(3P_2 ,n)=0.1$.
%%, as motivated by the result of  \cite{SF03}.
The $T_{\rm s} - T_{\rm in}$ relation is given by ``our fit''
curve of Fig. \ref{T-in}. Here and in all subsequent figures the
value $T_{\rm s}$ is the red-shifted temperature. NS masses are
indicated in the legend. For more details see  \cite{BGV04}.
\label{fig21BGV}}
\end{figure}
%%%%%%%%%%%%%%%%%%%%%%%%%%%%%%%%%%%%%%%%%%%%%%%%

Contrary to expectations of \cite{SF03} a more recent
work of \cite{KCTZ04} argued that the
%%$^3$P$_2$
$3P_2$ neutron pairing gap should be
dramatically enhanced,
as the consequence of  the strong softening of the pion propagator. According
to their estimation,
the
%%$^3$P$_2$
$3P_2$ neutron pairing gap is  as large as $1\div 10$~MeV in a
broad region of densities, see Fig. 1 of their work. Thus results
of calculations of \cite{SF03} and \cite{KCTZ04}, which both had
the same aim to include medium effects in the evaluation of the
$3P_2$ neutron gaps, are in a deep discrepancy with each other.

Note that in order to apply these results to a broad density
interval both models may need further improvements. The model  of
\cite{SF03} was developed to describe not too high densities. It
does not incorporate higher order nucleon-nucleon hole loops and
the $\Delta$ isobar contributions and thus it may only partially
include the pion softening effect at densities $\gsim n_0$.
Contrary, the model of \cite{KCTZ04} uses a  simplified analytic
expression for the effective pion gap $\widetilde{\omega}^2 (k_m)
=\mbox{min}_{k}[-D^{-1}(\omega =0,k=k_m )]$, where $D^{-1}$ is the
in-medium pion propagator,  valid near the pion condensation
critical point, if the latter occurred by a second order phase
transition. The latter assumption means that $\widetilde{\omega}^2
(k_m)$ is assumed to be zero in the critical point of the phase
transition. Outside the vicinity of the critical point the
parameterization  of the effective pion gap that was used can be
considered only as a rough interpolation. Actually the phase
transition is of first order and evaluations of quantum
fluctuations done by
 \cite{D82}
show that the value of the jump of the effective pion gap  in the
critical point is not as small. Moreover  repulsive  correlation
contributions to the $NN$ amplitude have been disregarded. In the
pairing channel under consideration, already outside  a narrow
vicinity of the pion condensation critical point, the repulsion
originating from the $NN$ correlation effects may exceed the
attraction originating from  the pion softening. Notice that, if
the pairing gap enhancement occurred only in a rather narrow
vicinity of the pion condensation critical point, it would not
affect the results of \cite{BGV04}. In the latter work two
possibilities were considered: i) a saturation of the pion
softening with increase of the baryon density resulting in the
absence of the pion condensation and ii) a stronger pion softening
stimulating the occurrence of the pion condensation for $n>n_c
\simeq 3~n_0$. In both cases the effective pion gap was assumed
never approaching zero and undergoing a not too small jump at the
critical point from a finite positive value ($\widetilde{\omega}^2
\simeq 0.3~m_{\pi}^2$, $m_{\pi}$ is the pion mass) to a finite
negative value ($\widetilde{\omega}^2 \simeq -0.1~m_{\pi}^2$). The
reason for such a strong jump is a strong coupling. If it were so,
a strong softening assumed by \cite{KCTZ04} would not be realized.
However, due to uncertainties in the knowledge of forces acting in
strong interacting nuclear matter and a poor description of the
vicinity of the phase transition point we can't exclude that the
alternative possibility of a tiny jump of the pion gap exists.
Therefore we will check how these  alternative hypotheses may work
within our "nuclear medium cooling" scenario. Thus avoiding
further discussion of the theoretical background of the models, in
this note we investigate the possibility of a significantly
enhanced
%%$^3$P$_2$
$3P_2$ neutron
pairing gap and of a partially suppressed proton $1S_0$ gap, as it has been
suggested by  \cite{KCTZ04}. To proceed  in the framework of our
 ``nuclear medium cooling'' scenario
we introduce the enhancement factor of the original $3P_2$ neutron
pairing gap $f (3P_2 ,n)$, and a suppression factor of the proton $1S_0$
gap $f(1S_0 ,p)$.  We do not change the neutron $1S_0$ gap since in
any case its effect on the cooling is minor.

 Generally speaking, the
suppression factors of superfluid processes are given by
complicated integrals. As it was demonstrated  by \cite{Sedr04} on
the example of the DU process, these integrals are, actually, not
reduced to the so called $R$-factors, see \cite{YGKLP03}. However,
for temperatures essentially below the critical temperature the
problem is simplified. With an exponential accuracy the
suppression of the specific heat is governed by  the factor
$\xi_{nn}$ for neutrons, and $\xi_{pp}$ for protons:
\begin{eqnarray}
\xi_{ii}\simeq  \mbox{exp}[{-\Delta_{ii}(T)/T}], \quad \mbox{for}\quad T<T_{ci};\quad i=n,p,
\end{eqnarray}
and $\xi_{ii}=1$ for $T>T_{ci}$, $T_{ci}$ is the corresponding critical
temperature.
We do not need a higher accuracy to
demonstrate our result. Therefore we will
use these simplified factors.

For the emissivity of the DU process the suppression factor is given by
$\mbox{min}\{\xi_{nn},\xi_{pp}\}$, see \cite{LP}.
Suppression factors for two nucleon processes follow from this fact and from
the diagrammatic representation of different processes within
the closed diagram technique by \cite{VS87}. These are:
 $\xi_{nn}\cdot \mbox{min}\{\xi_{nn},\xi_{pp}\}$
for the neutron branch of the  MU  process (and for the medium
modified Urca process, MMU); $\xi_{pp}\cdot
\mbox{min}\{\xi_{nn},\xi_{pp}\}$ for the  corresponding proton
branch of the process; $\xi_{nn}^2$ for the  neutron branch of the
(medium modified) nucleon bremsstrahlung (MnB) and $\xi_{pp}^2$
for the corresponding  proton branch of the bremsstrahlung (MpB).
Thus, for $\Delta_{nn}\gg \Delta_{pp}$ both neutron and proton
branches of the MMU process  are frozen for $T\ll T_{cn}$  due to
the factors $\xi_{nn}^2$ and $\xi_{pp}\xi_{nn}$, respectively.

The resulting cooling curves depend on the $T_{\rm in}-T_{\rm s}$
relation between internal and surface temperatures in the
envelope. Fig. \ref{T-in} shows uncertainties existing in this
relation. Calculation is presented for  the canonical NS: $M=1.4
M_{\odot}$, $R=10$~km with the crust model HZ90 of \cite{YLPGC03}.
Below we will show that a minimal discrepancy with the data is
obtained with ``our fit'' model. Using other choices like
``Tsuruta law" ($T_{\rm s}^{\rm Tsur}=(10 T_{\rm in})^{2/3}$,
where $T_{\rm s}$ and $T_{\rm in}$ are measured in K) only
increases the discrepancy. To compare results with ``our fit''
model we  use the upper boundary curve, ``$\eta =4\cdot
10^{-16}$\,'' and the lower boundary curve ``$\eta =4\cdot
10^{-8}$\,''. In Fig.\ref{T-in} we also draw lines $\eta
  =1\cdot 10^{-14}$ and $\eta =1\cdot 10^{-11}$ as they are
  indicated in the corresponding Fig. 2 of \cite{YLPGC03}.
In reality the selection of $\eta =4\cdot 10^{-8}$
and $\eta =4\cdot 10^{-16}$ as the boundaries of the
uncertainty-band seems to be a too strong restriction, see
\cite{YLPGC03}. The limit of the most massive helium layer is
achieved for
 $\eta \sim
10^{-10}$.
On the other hand the
 helium layer begins to affect the thermal structure
only for $\eta >
10^{-13}$. Thus one could exploit $10^{-13}<\eta < 10^{-10}$, as a $T_{\rm
  in}-T_{\rm s}$ band.
  We will use a broader band, as it is shown in
Fig.\ref{T-in}. By this we simulate effect of maximum
uncertainties in the knowledge of the $T_{\rm  in}-T_{\rm s}$
relation.

We present Fig. 21 of \cite{BGV04}, now Fig. \ref{fig21BGV}.
Cooling curves shown in
 this figure were calculated using
 ``our fit'' model of the crust, demonstrated by
the solid curve in Fig. \ref{T-in}. Here and in the corresponding
figures below the surface temperature is assumed to be
red-shifted, as it is inferred by the observer from the radiation
spectrum. Gaps are given by the model II of Fig. \ref{fig-gaps}.
However, the
%%$^3$P$_2$
$3P_2$ gap is additionally suppressed by a factor $f (3P_2
,n)=0.1$, as motivated by calculations of \cite{SF03}. If we took
the original
%%$^3$P$_2$
$3P_2$ gap of the model II, we would not succeed to describe the
data. The cooling then would be too fast, see Fig. 22 of
\cite{BGV04}. Now we will check the possibility of  ultra-high
%%$^3$P$_2$
$3P_2$
neutron pairing gaps, as motivated by  \cite{KCTZ04}.

{\em{In case when neutron processes are frozen the most
efficient process is the pPBF process}}, $p\rightarrow
p\nu\bar{\nu}$, for $T<T_{cp}$. Taking into account medium effects in the weak
coupling vertex we use the same expression for the emissivity of this process
as has been used by \cite{V00,BGV04}:
\begin{eqnarray}\label{PFB}
&&\varepsilon_{\nu} [\mbox{(M)pPBF}]\sim 10^{29}~\frac{m_N^*}{m_N}
\left[\frac{p_{{\rm F},p}}{p_{{\rm F},n}(n_0)}\right]
~\left[\frac{\Delta_{pp}}{\mbox{MeV}}\right]^{7}~
\nonumber\\
&&\times \left[\frac{T}{\Delta_{pp} }\right]^{1/2}~\xi_{pp}^2 ~~\frac{{\rm
    erg}}{{\rm cm}^{3}~{\rm sec}}  ~, \quad T<T_{cp}.
\end{eqnarray}
We point out that this process contributes only below the critical
temperature for the proton pairing. Inclusion of medium effects
greatly enhances the vertex of this process compared to the vacuum
vertex. Due to that a factor $\sim 10^2$ arises, since the process
may occur through $nn^{-1}$ and $ee^{-1}$ correlation states
(${-1}$ symbolizes the particle hole), with subsequent production
of $\nu\bar{\nu}$ from the $nn^{-1}\nu\bar{\nu}$ and
$ee^{-1}\nu\bar{\nu}$ channels rather than from a strongly
suppressed channel $pp^{-1}\nu\bar{\nu}$, see
\cite{VS87,SV87,MSTV90,VKK98,L00,V00}. Relativistic corrections
incorporated in the description of the $pp^{-1}\nu\bar{\nu}$
vertex  also produce an enhancement but quite not as strong as
that arising from medium effects in $nn^{-1}\nu\bar{\nu}$ and
$ee^{-1}\nu\bar{\nu}$ channels. We point out that we see no
arguments not to include these corrections and we pay attention to
only a moderate dependence  of the result on the uncertainties in
the knowledge of the strong interaction.

We also present here an explicit expression for the emissivity of
the proton branch of the nucleon bremsstrahlung including medium
effects, MpB, $pp\rightarrow pp\nu\bar{\nu}$. In case of
suppressed neutron $3P_2$ gaps this process contributed much less
than several others. However, in case when neutron processes are
frozen, the $pp\rightarrow pp\nu\bar{\nu}$ process becomes the
dominating process for $T_{cn}>T>T_{cp}$. The emissivity of the
$pp\rightarrow pp\nu\bar{\nu}$ reaction takes the form (see
\cite{VS86} for more details)
\begin{eqnarray}\label{MpB}
&&\epsilon({\rm MpB})\sim  10^{23}\xi_{pp}^2  I_{pp}\frac{Y_p^{5/3}
 \Gamma_w^2 \Gamma_s^4 T_9^8}{\widetilde{\omega}^4 }\nonumber\\
&&\times\left(\frac{m_N^*}{m_N}\right)^4
\left(\frac{n}{n_0}\right)^{5/3}~~\frac{{\rm
    erg}}{{\rm cm}^{3}~{\rm sec}},
\end{eqnarray}
%%in $\mbox{erg}/(\mbox{cm}^3\cdot \mbox{sec})$,
$T_9 =T/10^9$ K, $m_N^* $ is the effective nucleon
 mass,  $\Gamma_w \simeq 1$, and
 $\Gamma_s \simeq 1/[1+C(n/n_0)^{1/3}]$, $C\simeq 1.4\div 1.6$, take into
 account $NN$ correlations
in weak and strong interaction vertices, respectively,
 $Y_p =n_p/n$ is the proton to nucleon ratio.
%%$\widetilde{\omega}^2 (k_m)$ is the effective pion gap minimized in the transverse momentum.
We for simplicity  assumed that the value $k=k_m$, at which the
effective pion gap $\widetilde{\omega}^2 (k)$ gets the minimum, is
rather close to the value of the neutron Fermi momentum $ p_{{\rm
F},n}$
 (as it follows from the microscopic analysis
of \cite{MSTV90}).
To simplify the consideration we take the same value of the effective
pion gap for the given process  as that for the MMU
 process (although in general case it is not so, and thus the result
 (\ref{MpB}) proves to be essentially model dependent),
cf. \cite{BGV04},
\begin{eqnarray}\label{intI}
I_{pp} \sim \frac{\pi}{64}\left(\frac{p_{{\rm F},n}}{p_{{\rm F},p}}\right)^5 \frac{\widetilde{\omega}}{p_{{\rm F},n}}.
\end{eqnarray}
We have checked that for $T<T_{cp}$ for the pairing gaps under
consideration the MpB reaction contributes significantly less than
the pPBF process. It could be not the case only  in a narrow
vicinity of the pion condensation critical point, if pion
condensation occurred with only a tiny jump of the effective pion
gap in the  critical point. However, even in this case there are
many effects which could mask this abnormal enhancement.

In case of frozen neutron degrees of freedom the specific heat is governed by protons and electrons:
\begin{equation}
c_p \sim 10^{20}({m_N^*}/{m_N})~(n_p/n_0)^{1/3} \xi_{pp}~T_9 ~
{\rm erg~cm^{-3}K^{-1}}~,
\end{equation}
\begin{equation}\label{e}
c_e \sim 6\times 10^{19} \,(n_e/n_0)^{2/3}~T_9 ~
{\rm erg~cm^{-3}~K^{-1}}~.
\end{equation}
Here, we again suppress a contribution to the specific heat of a
narrow vicinity of the pion condensation critical point  due to
the fact that in our scenario (see Fig. 1 of  \cite{BGV04}) the
modulus of the effective pion gap $\widetilde{\omega}^2 $ is
always larger than $\sim (0.1\div 0.3)~m_{\pi}^2$. With such an
effective pion gap the pion contribution to the specific heat  is
not too strong and can be disregarded in order to simplify the
consideration. For the second order phase transition (either for a
first order phase transition but with a tiny jump of
$|\widetilde{\omega}^2 |$ in the critical point), pion
fluctuations would contribute  stronger to the specific heat
yielding a term $c_{\pi} \propto T/\widetilde{\omega}$, see
\cite{VM82,MSTV90}.

In Figs. \ref{fig50,0.1} and \ref{fig50,0.5} we demonstrate the
sensitivity of the results presented in Fig. \ref{fig21BGV} to the
enhancement of the neutron $3P_2$ gap and to a suppression of the
$1S_0$ proton gap, following the suggestion of \cite{KCTZ04}. We
start with  the ``our crust'' model and the model II for the gaps,
using however the additional
%%We will present results of calculations with
enhancement factor $f(3P_2, n)=50$ for the neutron $3P_2$ gap.
Introducing factors $f(1S_0 ,p)=0.1$ and $f(1S_0 ,p)=0.5$ we test
the sensitivity of the results to the variation of the $1S_0$
proton gap. We do not change the value of the $1S_0$ neutron gap
since its variation almost does not influence on the cooling
curves for NS's with masses $M>1~M_{\odot}$, that we will
consider.

Comparison of Figs. \ref{fig21BGV}
%%and Figs. \ref{fig10,0.1}
 -- \ref{fig50,0.5}
shows that in all cases NS's with masses $M\gsim 1.8~M_{\odot}$
cool similar in spite of the fact that $3P_2$ neutron and $1S_0$
proton gaps are varied in wide limits. This is because $3P_2$
neutron and $1S_0$ proton gaps disappear at the high densities,
being achieved in the central regions of these very massive NS's,
see Fig. \ref{fig-gaps}. Thus these objects cool down similar to
non-superfluid objects. Extremely rapid cooling of stars with
$M\geq 1.84~M_{\odot}$ is due to the DU process, being very
efficient in the normal matter. Thereby we will notice that
{\em{the cooling curves are very sensitive to  the density
dependence of the gaps.}}
%%%%%From comparison of Figs. \ref{fig50,0.1} and  \ref{fig50,0.5} we see  that
The difference in the cooling of NS's with $M<1.8~M_{\odot}$ in
cases presented by Figs. \ref{fig50,0.1} and  \ref{fig50,0.5} is
the consequence of  different values of proton gaps used in these
two calculations. This difference is mainly  due to the pPBF
processes. The larger the proton gap, the higher is the
emissivity.
%%For that we use an
%%enhancement factor $f (3P_2 ,n)$.

We checked that for stars with $M\lsim 1.6~M_{\odot}$ for $T<T_{cn}$ for the
%%$^3$P$_2$
$3P_2$ neutron pairing, a complete freezing of neutron degrees of
freedom occurs already for $f (3P_2 ,n)\gsim 20$.  Contributions
to the emissivity and to the specific heat  involving neutrons are
fully suppressed then. For heavier stars ($M>1.6~M_{\odot}$) a
weak dependence on the value of the factor $f(3P_2 ,n)$ still
remains even for $f(3P_2 ,n)>100$ but the corresponding cooling
curves lie too low to allow for an appropriate fit of the data.
This difference between cooling of stars with $M<1.6~M_{\odot}$
and $M>(1.6\div 1.7)~M_{\odot}$ is due to  the mentioned density
dependence of the neutron $3P_2$ gap. The latter value smoothly
decreases with increase of the density reaching zero for $n\gsim
4.5~n_0$ (the density $4.5~n_0$ is achieved in the center of a NS
of the mass $M = 1.7~M_{\odot}$). At densities slightly below
$4.5~n_0$ the gap is rather small. Therefore for stars with
$M>(1.6\div 1.7)~M_{\odot}$ the scaling of the gap by  a factor
$f(3P_2 ,n)$ changes the size of the region where  the gaps may
affect the cooling. For stars with $M\lsim 1.6~M_{\odot}$ gaps
have finite values even at the center of the star. Thereby there
exists a critical value of the factor $f(3P_2 ,n)$, such that for
higher values of $f(3P_2 ,n)$ the cooling curves are already
unaffected by its change.

Figs. \ref{fig50,0.1} and  \ref{fig50,0.5} demonstrate that
we did not
succeed to reach appropriate overall agreement with the data getting too rapid
cooling.
 If we used the effective pion gap that
allowed for the second order pion condensation transition, as \cite{KCTZ04} assumed, we would get much
more rapid cooling that would  disagree even more with the data.
The cooling of the old pulsars is
not explained in all cases. Although the heating mechanism used by \cite{T04} may partially
help in this respect,
the discrepancy between the curves and the data points seems to be too high,
especially in Fig. \ref{fig50,0.5}.
We see that in our regime of
frozen neutron processes a better
fit is achieved in Fig.
%%%%\ref{fig10,0.1} and
\ref{fig50,0.1}, i.e.,  for a stronger suppressed proton gap (for
$f(1S_0 ,p)=0.1$). Actually we note that the discrepancy is even
more severe, since to justify the idea of  \cite{KCTZ04} we should
exploit a softer pion propagator. Only a strong softening of the
pion mode might be consistent with significant increase of the
neutron $3P_2$ gap. On the other hand such an additional softening
would immediately result in a still more rapid cooling. The work
of \cite{VKZC} discussed the possibility of a novel very efficient
process with the emissivity $\epsilon_{\nu}\propto T^5$, that
would occur due to non-fermi liquid behavior of the Fermi sea in a
narrow  vicinity of the pion condensation critical point at the
assumption of a strong pion softening. If we included this very
efficient process, the disagreement with the data could be
strongly enhanced. An enhancement of the specific heat due to pion
fluctuations within the same vicinity region of the pion
condensation point can't compensate the acceleration of the
cooling owing to the enhancement of the emissivity. \cite{KCTZ04}
used the value $n_{c}=2~n_0$ for the critical density of the pion
condensation. In case of the Urbana-Argonne equation of state that
we exploit here (we use the HHJ fit of this equation of state that
removes the causality problem, see \cite{BGV04} for details) the
density $n=2~n_0$ is achieved in the central region of a NS with
the mass $M\sim 0.8~M_{\odot}$. This means that all NS's with $M
\gsim 0.8~M_{\odot}$ would cool extremely fast and would not be
seen in soft $X$ rays.

\begin{figure}[ht]
%%\vspace{-0.5cm}
\centerline{
\psfig{figure=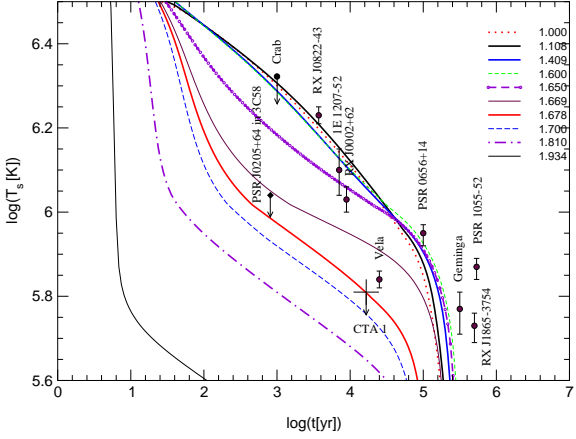,height=0.4\textwidth,angle=-90}}
\caption{Cooling curves according to the nuclear medium cooling
scenario, see Fig. \ref{fig21BGV}. Gaps are from Fig.
\ref{fig-gaps} for model II but the $3P_2$ neutron
  pairing gap is additionally enhanced  by a factor $f(3P_2 ,n)=50$ and  the
  $1S_0$ proton gap is suppressed by $f(1S_0 ,p)=0.1$.
The $T_{\rm s} - T_{\rm in}$ relation is given by ``our fit''
curve of Fig. \ref{T-in}. } \label{fig50,0.1}
\end{figure}

\begin{figure}[ht]
%%\vspace{-0.5cm}
\centerline{
\psfig{figure=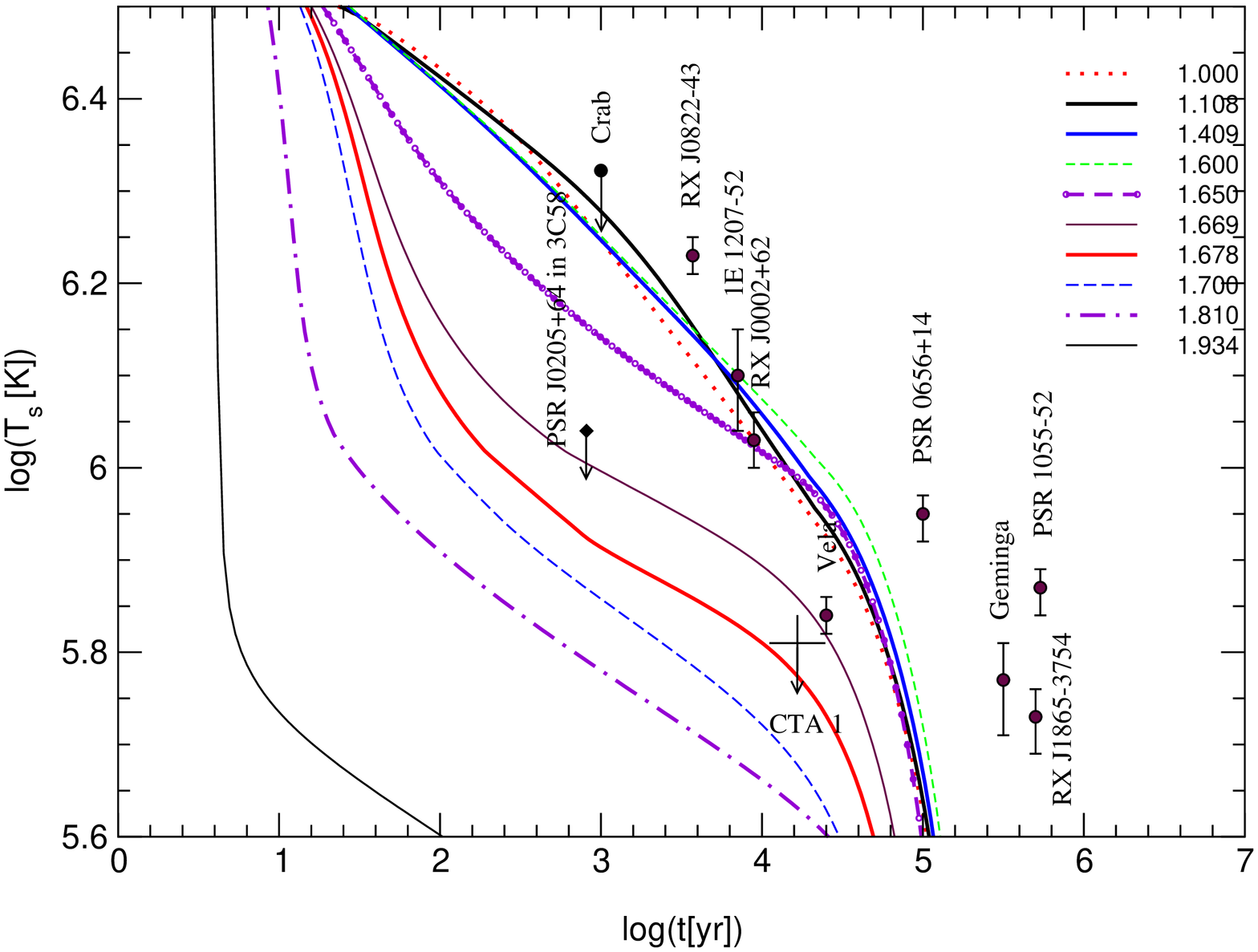,height=0.4\textwidth,angle=-90}}
\caption{Cooling curves according to the nuclear medium cooling
scenario, see Fig. \ref{fig21BGV}. Gaps are from Fig.
\ref{fig-gaps} for model II but the $3P_2$ neutron
  pairing gap is additionally enhanced  by a factor $f(3P_2 ,n)=50$ and  the
  $1S_0$ proton gap is suppressed by $f(1S_0 ,p)=0.5$.
The $T_{\rm s} - T_{\rm in}$ relation is given by ``our fit''
curve of Fig. \ref{T-in}; see also
%%Fig. \ref{fig10,0.1}.Same as in
Fig. \ref{fig50,0.1}.
%% but for the original $1S_0$ proton gap suppressed by $f(1S_0 ,p)=0.5$.
%%%%{\bf{Hovik please
%%%%     replace this fig. to gap =50}}
} \label{fig50,0.5}
\end{figure}

\begin{figure}[ht]
%%\vspace{-0.5cm}
\centerline{
\psfig{figure=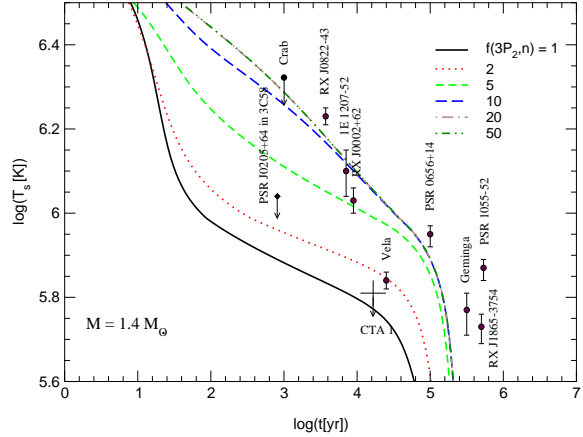,height=0.4\textwidth,angle=-90}}
\caption{Cooling curves of the neutron star with the mass
$1.4~M_{\odot}$ according to the nuclear medium cooling scenario,
see Fig. \ref{fig21BGV}. Gaps are from Fig. \ref{fig-gaps} for
model II but the $3P_2$ neutron
  pairing gap is additionally enhanced  by different factors $f(3P_2 ,n)$
  (shown in Figure)
and  the
  $1S_0$ proton gap is suppressed by $f(1S_0 ,p)=0.1$.
The $T_{\rm s} - T_{\rm in}$ relation is given by ``our fit''
curve of Fig. \ref{T-in}. } \label{figM1.4}
\end{figure}

Actually, we checked the whole interval of variation of $f(3P_2,
n)$ and $f(1S_0 ,p)$ factors in the range $1\div 100$ and $0.1\div
0.5$ respectively. We verified that the variation of $f(3P_2, n)$
and $f(1S_0 ,p)$ factors in the whole mentioned range done within
our parameterization of the effective pion gap does not allow to
improve the picture. In all cases we obtain too fast cooling. To
demonstrate this in Fig.~\ref{figM1.4} we show the cooling of a
$1.4~M_{\odot}$ star for different values of the $f(3P_2, n)$
factor. The factor $f(1S_0 ,p)$ is taken to be 0.1. We see that
for $f(3P_2, n)<15\div 20$ the curves rise with the increase of
$f(3P_2, n)$ factor. For $f(3P_2, n)>20$ the curves  do not depend
on $f(3P_2, n)$.

To check how the results are sensitive to uncertainties in our
knowledge of the value  (\ref{intI}) that determines the strength
of the in-medium effect on the emissivity of the MpB process we
multiplied (\ref{MpB}) by a pre-factor $f(\rm MpB)$  that we
varied in a range $f(\rm MpB)=0.2\div 5$. In agreement with the
above discussion, for $f(\rm MpB)<1$, for temperatures $\mbox{log}
T_{\rm s}[\rm K]>5.9$ the cooling curves are shifted upwards.
Opposite, for $f(\rm MpB)>1$, for temperatures $\mbox{log} T_{\rm
s}[\rm K]>5.9$ the cooling curves are shifted downwards. However
independently of the value $f(\rm MpB)$ for $\mbox{log} T_{\rm
s}[\rm K]<5.9$ curves are not changed. Thus it does not allow to
diminish the discrepancy with the data.

\begin{figure}[htb]
%\vspace{-0.5cm}
\centerline{
\psfig{figure=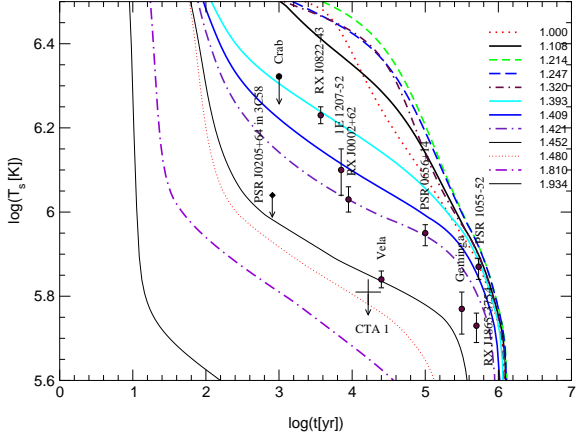,height=0.4\textwidth,angle=-90}}
\caption{Fig. 15 of \cite{BGV04}. Gaps from the model I. The
original $3P_2$  neutron gap is suppressed by $f(3P_2 ,n)=0.1$.
The $T_{\rm s} - T_{\rm in}$ relation is given by ``our fit''
curve of Fig. \ref{T-in}. Other notations
  see in \cite{BGV04}.} \label{fig15BGV}
\end{figure}

Now we will check the efficiency of another choice of the gaps, as
motivated by the model I, thick lines in Fig. \ref{fig-gaps}.
Compared to the model II the model I uses an {\em{artificially
enhanced proton gap}}. Thereby, one can expect that the model I is
less realistic than the model II. Also we pay attention to a
different density dependence of the proton gap (it cuts off for
densities $n\gsim 3n_0$ in the model I) compared to that given by
the model II. However, as we have mentioned, uncertainties in
existing calculations of the gaps are very high. Thus  it is
worthwhile  to check different possibilities. Since the mentioned
parameterization has been used by one of the groups working on the
problem of cooling of NS's, see \cite{YGKLP03}, we will  consider
consequences of this possibility as well. Fig. \ref{fig15BGV}
demonstrates our previous  fit of the data  within the model I,
but for the original $3P_2$  neutron gap being suppressed by
$f(3P_2 ,n)=0.1$ (see \cite{BGV04}).

Figs. \ref{fig50,0.1Y} and \ref{fig50,0.5Y} show that within the
variation of the gaps of the model I the discrepancy  with the data is still
stronger compared to that for the above calculation  based on the use of the model II.

The difference between curves shown in Figs. \ref{fig50,0.1Y} and
\ref{fig50,0.5Y} is less pronounced than for those curves
demonstrated in Figs.  \ref{fig50,0.1} and  \ref{fig50,0.5}.
Indeed,  as we have mentioned, the density dependence of the
proton gap is different in models I and II. In the model II  the
proton gap reaches up to higher densities than in the model I (in
the latter case the gap is cut already  for $n\gsim 3n_0$). Thus
in the  case shown by Figs. \ref{fig50,0.1Y} and  \ref{fig50,0.5Y}
a non-superfluid core begins to contribute already for smaller
values of the star mass.
%%for stars with $M\gsim 1.3~M_{\odot}$  essentially contributes to the cooling
%%Figs. \ref{fig50,0.1Y} and  \ref{fig50,0.5Y} demonstrate
Therefore  in both figures the corresponding cooling curves are almost the same
for  $M\gsim 1.6 M_{\odot}$.

\begin{figure}[ht]
%%\vspace{-0.5cm}
\centerline{
\psfig{figure=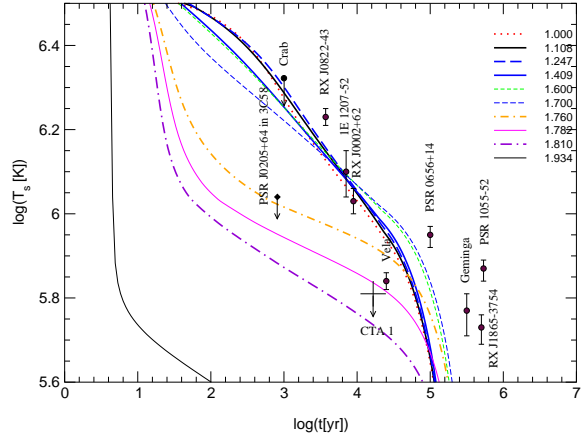,height=0.4\textwidth,angle=-90}}
\caption{Cooling curves according to the nuclear medium cooling
scenario, see Fig. \ref{fig15BGV}. Gaps are from Fig.
\ref{fig-gaps} for model I but the $3P_2$ neutron
  pairing gap is additionally enhanced  by a factor $f(3P_2 ,n)=50$ and  the
  $1S_0$ proton gap is suppressed by $f(1S_0 ,p)=0.1$.
The $T_{\rm s} - T_{\rm in}$ relation is given by ``our fit''
curve of Fig. \ref{T-in}. } \label{fig50,0.1Y}
\end{figure}
\begin{figure}[ht]
%%\vspace{-0.5cm}
\centerline{
\psfig{figure=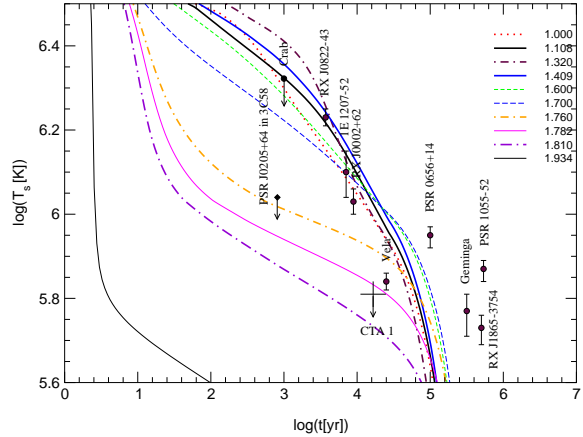,height=0.4\textwidth,angle=-90}}
\caption{Same as Fig. \ref{fig50,0.1Y}, but for the original
 $1S_0$ proton gap suppressed by $f(1S_0 ,p)=0.5$.
} \label{fig50,0.5Y}
\end{figure}

The dependence of the results on the different choices of the
$T_{\rm s} - T_{\rm in}$ relation is demonstrated  by Figs.
\ref{fig10,0.1etII}, \ref{fig10,0.5etII} for gaps based on a
modification of the model II and by Figs.  \ref{fig10,0.1etI},
\ref{fig10,0.5etI} for gaps based on a modification of the model
I.  For this demonstration we first took the upper boundary curve
$\eta =4\cdot 10^{-16}$ in Fig.~\ref{T-in}. We show that these
choices however do not allow to improve the fit. Comparing Figs.
\ref{fig10,0.1etII} and  \ref{fig50,0.1} (and \ref{fig10,0.5etII}
and \ref{fig50,0.5}) based on the very same modification of the
model II we see that with the ``our fit'' crust model the
deviation from the data points is less pronounced. Basing on the
model I we compare Figs. \ref{fig10,0.1etI} and  \ref{fig50,0.1Y}
(and \ref{fig10,0.5etI} and \ref{fig50,0.5Y}) and we arrive at the
very same conclusion.

\begin{figure}[ht]
\vspace{-0.5cm} \centerline{
\psfig{figure=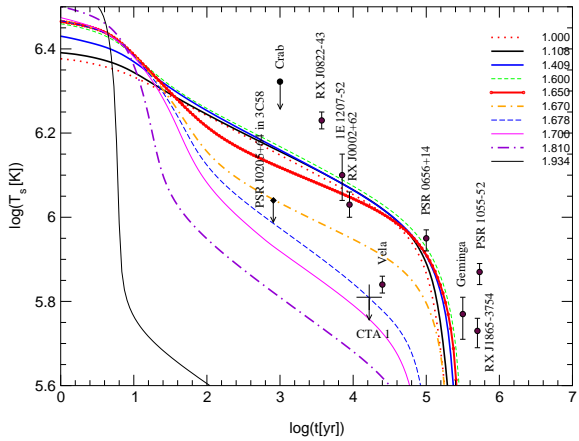,height=0.4\textwidth,angle=-90}}
\caption{Same as in Fig. \ref{fig50,0.1}, but for the crust
model $\eta
  =4.0 \cdot 10^{-16}$.
} \label{fig10,0.1etII}
\end{figure}

\begin{figure}[ht]
\vspace{+0.5cm} \centerline{
\psfig{figure=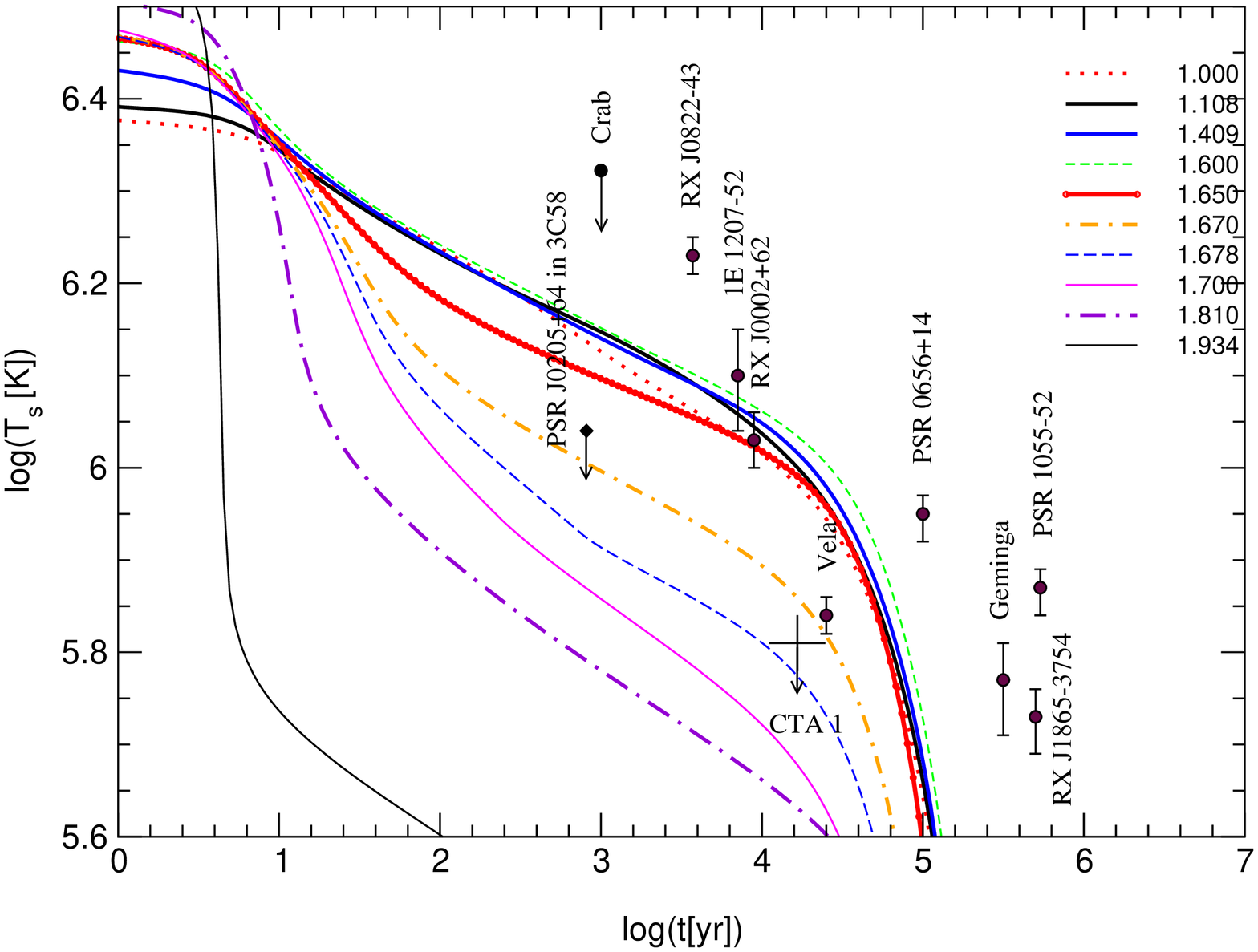,height=0.4\textwidth,angle=-90}}
\caption{Same as in Fig.\ref{fig50,0.5}, but for the crust
model $\eta =4.0 \cdot 10^{-16}$. }
\label{fig10,0.5etII}
\end{figure}

\begin{figure}[ht]
\vspace{-0.5cm} \centerline{
\psfig{figure=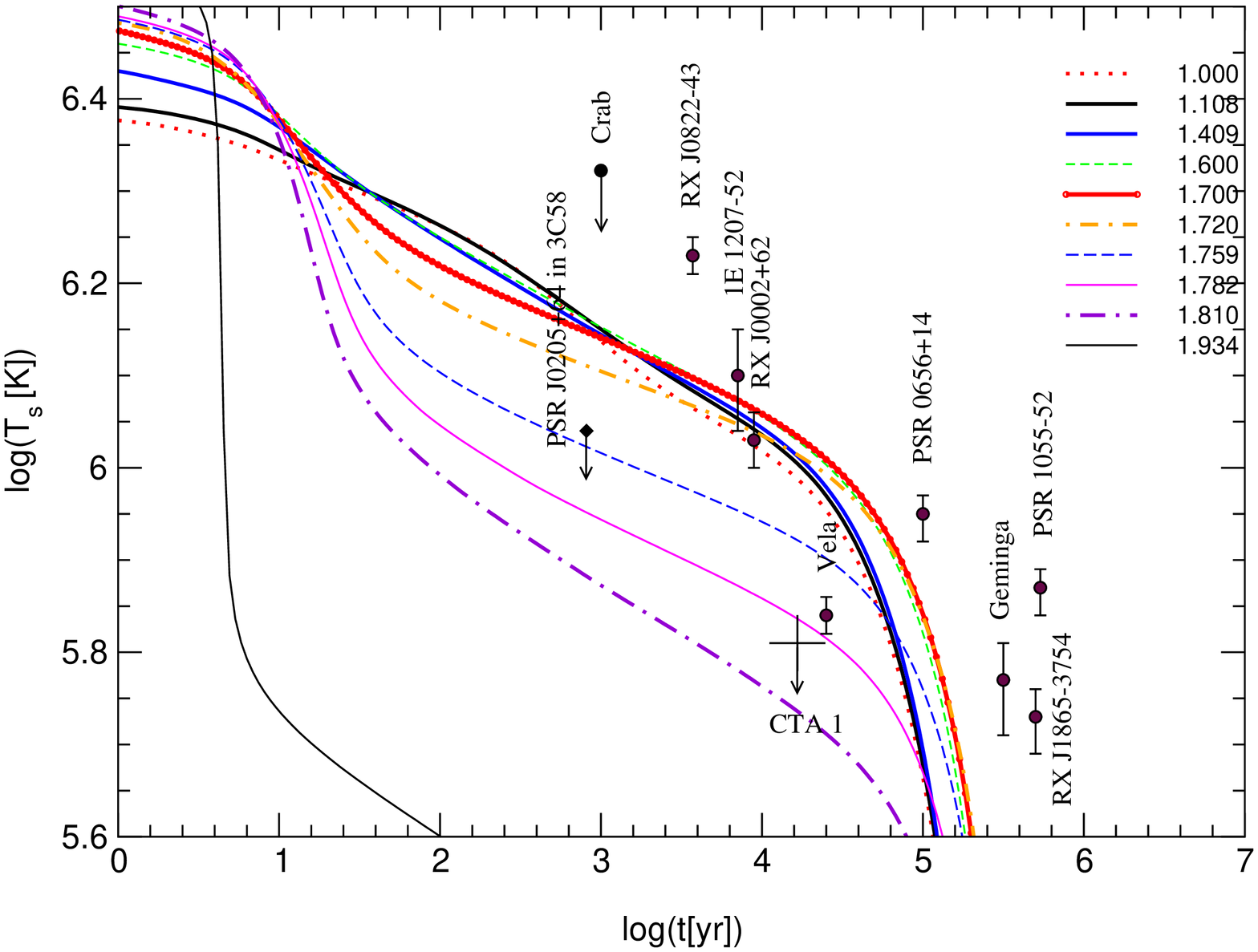,height=0.4\textwidth,angle=-90}}
\caption{Same as in Fig.\ref{fig50,0.1Y}, but for the crust
model $\eta =4.0 \cdot 10^{-16}$. }
\label{fig10,0.1etI}
\end{figure}

\begin{figure}[ht]
\vspace{+0.5cm} \centerline{
\psfig{figure=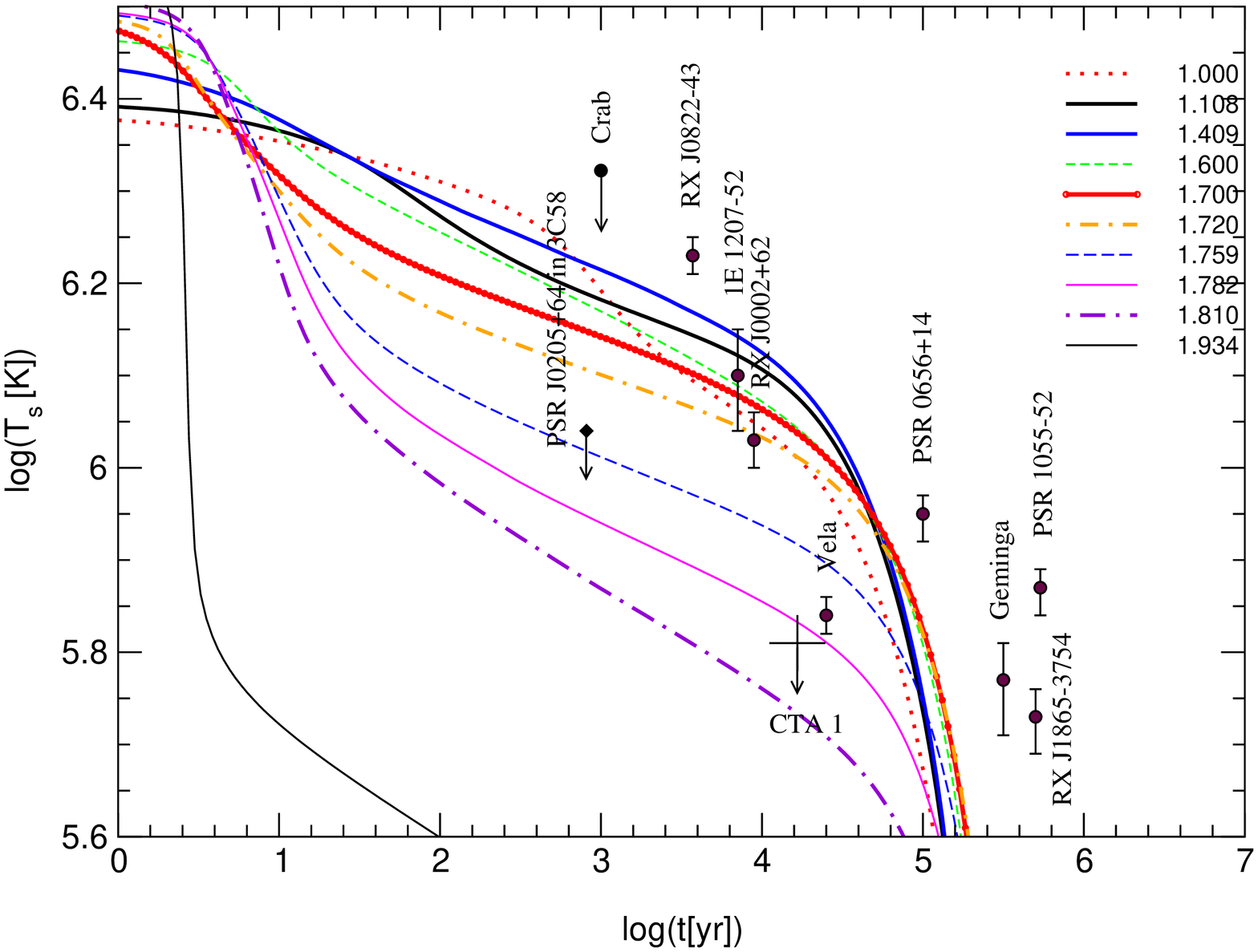,height=0.4\textwidth,angle=-90}}
\caption{Same as in Fig. \ref{fig50,0.5Y}, for the crust model
$\eta =4.0 \cdot 10^{-16}$. }
\label{fig10,0.5etI}
\end{figure}

\begin{figure}[ht]
%%\vspace{-0.5cm}
\centerline{
\psfig{figure=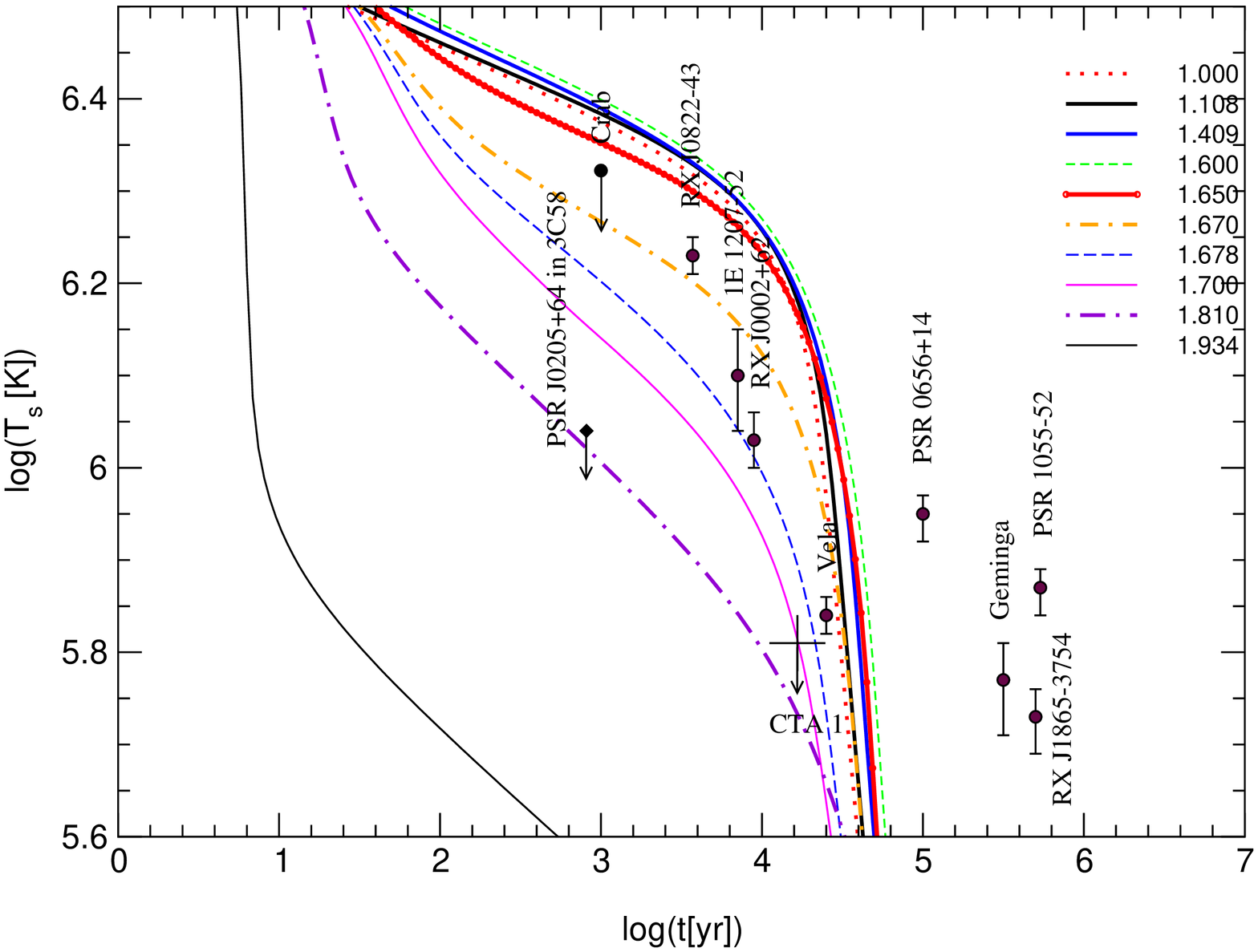,height=0.4\textwidth,angle=-90}}
\caption{Same as in Fig. \ref{fig50,0.1}, but for the crust
model $\eta
  =4.0 \cdot 10^{-8}$.
} \label{fig50,0.1cr}
\end{figure}

In Fig. \ref{fig50,0.1cr} we use the lower boundary curve $\eta
  =4.0 \cdot 10^{-8}$ of the Fig. \ref{T-in}.
We further demonstrate that the selection of a different choice of
the $T_{\rm s}-T_{\rm in}$ relation within the band shown in
Fig.~\ref{T-in} does not allow to diminish discrepancy with the
data. Contrary, this
  discrepancy just increases compared to that demonstrated by   ``our fit''
  model.  Indeed, the cooling evolution for times $t\lsim 10^5$~yr
($T_{\rm s}\gsim
  10^{6}$K) is governed by neutrino processes. Thus the higher $T_{\rm in}$,
the larger $T_{\rm s}$ is. The slowest cooling is then  obtained, if one
  uses the lower boundary curve $\eta
  =4.0 \cdot 10^{-8}$ of Fig.~\ref{T-in}.
The evolution of NS's for   times $t\gsim 10^5$~yr begins to be controlled by
  the photon processes. In the photon epoch ($t\gg 10^5$~yr) the smaller the  $T_{\rm s}$ value, the
  less efficient the radiation is. Thus for $t\gg 10^5$~yr
the slowest cooling is obtained, if one uses the upper boundary curve $\eta
  =4.0 \cdot 10^{-16}$ of Fig.~\ref{T-in}. The ``our crust'' curve just simulates the
  transition from the one limiting curve to the other
demonstrating the slowest cooling in the whole temperature
interval shown in the figures.

We point out that in all cases the data are not explained within
the assumption of an enhanced neutron $3P_2$ gap (for $f(3P_2 ,n)
>1$) and of a partially suppressed $1S_0$ proton gap (for $f(1S_0
,p)=0.1\div 0.5$).

Our aim was to consider the possibility of large $3P_2$ gaps
within the same "Nuclear medium cooling" scenario of \cite{BGV04}
that well described the cooling data in opposite assumption of
suppressed $3P_2$ gaps. Therefore in the present work we did not
incorporate possibilities of internal heating for old pulsars (see
\cite{T04}) and of existence of quark cores (see
\cite{BGVquark1,BGVquark2} and refs therein).

The main problem with the given scenario is the following.   At
the frozen neutron contribution to the specific heat and to the
emissivity the region of surface temperatures $T_{\rm s} > 10^6$K
is determined by proton processes. The most efficient among them
is the pPBF process. For proton gaps, which we deal with, the pPBF
process proves to be too efficient yielding too rapid cooling.
Thus at least several slow cooling data points (at least data for
old pulsars) are not explained. Note that some works ignore the
mentioned above medium induced enhancement of the pPBF emissivity
that results in  10-100 times suppression of the rate. We omitted
this possibility as not physical one.

Concluding, we have shown that the ``Nuclear medium cooling''
scenario of \cite{BGV04} fails to appropriately fit the NS cooling
data at the assumption of a strong enhancement of the $3P_2$
neutron gaps (we checked the range  $f(3P_2 ,n)=1\div 100$) and
for moderately suppressed $1S_0$ proton gaps (for $f(1S_0
,p)=0.1\div 0.5$). On the other hand the very same scenario
allowed us to appropriately fit the whole set of data  at the
assumption of a significantly suppressed $3P_2$ neutron gap (for
$f(3P_2 ,n) \sim 0.1$). We observed an essential dependence of the
results not only on the values of the gaps but also on their
density dependence. We used the density dependence of the gaps
according to the models I and II. The latter model  is supported
by microscopic calculations. We excluded an attempt to
artificially fit  the density dependence of the gaps trying to
match cooling curves with the data. Although such an attempt could
improve the fit, this way seems us rather not physical and we did
not follow it. However we will  greatly  encourage further
attempts of microscopic  calculations of the gaps, which would
take into account most important medium effects. With carefully
treated gaps one could return to the simulation of the NS cooling.

We thank D. Blaschke for his permanent interest in our work,
critical reading of the manuscript and valuable remarks. We also
thank B. Friman and A. Schwenk for interesting discussions and
A.Y. Potechin for helpful comments. The research of H.G. was
supported by the Virtual Institute of the Helmholtz Association
under grant No. VH-VI-041 and by the DAAD partnership program
between the Universities of Rostock and Yerevan. The work of
D.N.V. was supported in part by the Deutsche
Forschungsgemeinschaft (DFG project 436 RUS 113/558/0-2) and the
Russian Foundation for Basic Research (RFBR grant 03-02-04008).


\begin{thebibliography}{}
\bibitem[Ainsworth et al. (1989)]{AWP}
Ainsworth, T., Wambach, J., Pines, D. 1989, Phys. Lett. B, {222},
173
\bibitem[Akmal et al. (1998)]{APR98}
Akmal, A., Pandharipande, V.R., Ravenhall, D.G. 1998, Phys. Rev. C, {58}, 1804
\bibitem[Blaschke et al. (2004)]{BGV04}
Blaschke, D., Grigorian, H., Voskresensky, D.N. 2004, A \& A,  {424}, 979
\bibitem[Blaschke et al. (2001)]{BGVquark1}
Blaschke, D., Grigorian, H., Voskresensky, D.N. 2001, A \& A,
{368}, 561
\bibitem[Dyugaev (1982)]{D82}
Dyugaev, A.M. 1982, JETP Lett. {35}, 420
\bibitem[Flowers et al. (1976)]{FRS76}
Flowers, E., Ruderman, M., Sutherland, P. 1976, ApJ, {205}, 541
\bibitem[Friman \& Maxwell (1979)] {FM79}
Friman, B., Maxwell, O.V. 1979, ApJ, {232}, 541
\bibitem[Grigorian et al. (2004)]{BGVquark2}
 Grigorian, H., Blaschke, D., Voskresensky, D.N.2004
 [arXiv:astro-ph/0411619]
\bibitem[Ivanov et al. (2001)]{IKHV01}
Ivanov, Yu.B., Knoll, J., van Hees, H., Voskresensky, D.N.
2001, Phys. Atom. Nucl., {64}, 652
\bibitem[Khodel et al. (2004)]{KCTZ04}
Khodel, V.A., Clark, J.W., Takano, M., Zverev, M.V. 2004, Phys.
Rev. Lett. 93, 151101
\bibitem[Lattimer et al. (1991)]{LP}
Lattimer, J.M., Pethick, C.J., Prakash, M.,  Haensel, P., 1991,
Phys. Rev. Lett. 66, 2701
\bibitem[Leinson (2000)]{L00}
Leinson, L.B. 2000,  Phys. Lett. B, {473}, 318
\bibitem[Lombardo \& Schulze (2000)]{LS00}
Lombardo, U., Schulze, H.-J. 2001, in: 'Physics of Neutron Stars
Interiors', Blaschke, D., Glendenning, N.K., Sedrakian, A. (Eds),
 Lect. Notes Phys. 578, 30
%[arXiv:astro-ph/0012209]
\bibitem[Migdal et al. (1990)]{MSTV90}
Migdal, A.B., Saperstein, E.E., Troitsky, M.A., Voskresensky, D.N.
1990, Phys. Rept., {192}, 179
\bibitem[Page et al. (2004)]{PLPS04}
Page, D., Lattimer, J.M., Prakash, M., Steiner, A.W. 2004,
ApJS, 155, 623
%[arXive: astro-ph/0403657]
\bibitem[Rapp  \& Wambach (1994)]{RW}
Rapp, R., Wambach, J. 1994, Nucl. Phys., A, 573, 626
\bibitem[Schaab et al. (1997)]{SVSWW97}
Schaab, Ch., Voskresensky, D., Sedrakian, A.D., Weber, F., Weigel,
M.K. 1997, A\& A, {321}, 591
\bibitem[Schulze et al. (1996)]{SCLBL96}
Schulze, H.-J., Cugnon, J., Lejeune, A., Baldo, M., Lombardo, U. 1996,
Phys. Lett. B, {375}, 1
\bibitem[Schwenk \& Friman (2004)]{SF03}
Schwenk, A., Friman, B. 2004, Phys. Rev. Lett. 92, 082501
%%; [arXiv: nucl-th/0307089]
\bibitem[Sedrakian (2005)]{Sedr04}
Sedrakian, A. 2005,  Phys. Lett. B, {607}, 27
\bibitem[Senatorov \& Voskresensky (1987)]{SV87}
Senatorov, A.V., Voskresensky, D.N. 1987, Phys. Lett. B, {184},
119
\bibitem[Shapiro \&  Teukolsky (1983)]{ST83}
Shapiro, S., Teukolsky, S.A. 1983, {\em{ Black Holes, White Dwarfs
and Neutron Stars: The Physics of Compact Objects},
Wiley-Interscience, New-York, USA}
\bibitem[Stairs (2004)]{S04}
Stairs, I.H. 2004, Science, {304}, 547
\bibitem[Takatsuka \& Tamagaki (2004)]{TT04}
Takatsuka, T., Tamagaki, R. 2004,  Prog. Theor. Phys. 112, 37
\bibitem[Tsuruta (1979)]{T79}
Tsuruta, S.  1979, Phys. Rept., 56, 237
\bibitem[Tsuruta (2004)]{T04}
Tsuruta, S. 2004, in: Proceedings of IAU Symposium
``Young Neutron Stars and their Environments'', F. Camilo, B.M. Gaensler
(Eds.), vol. 218,  [arXiv:astro-ph/0401245]
\bibitem[Tsuruta et al. (2002)]{TTTTT02}
Tsuruta, S., Teter, M.A., Takatsuka, T., Tatsumi, T., Tamagaki, R. 2002,
ApJ, {571}, L143
\bibitem[Voskresensky et al. (1998)]{VKK98}
Voskresensky, D.N., Kolomeitsev, E.E., K\"ampfer, B. 1998, JETP, {87}, 211
\bibitem[Voskresensky \& Senatorov (1984)]{VS84}
Voskresensky, D.N., Senatorov, A.V. 1984,
%Pis'ma v ZhETF. {\bf 40}, 395 (1984) (In Engl.:
JETP Lett., {\bf 40}, 1212
\bibitem[Voskresensky \& Senatorov (1986)]{VS86}
Voskresensky, D.N., Senatorov, A.V. 1986,
%ZhETF. {\bf 90}, 1505 (1986) (In Engl.:
JETP, {\bf 63}, 885
\bibitem[Voskresensky \& Senatorov (1987)]{VS87}
Voskresensky, D.N., Senatorov, A.V. 1987,
%Yad. Fiz. {\bf 45}, 657 (1987) (In Engl.:
Sov. J. Nucl. Phys., {45}, 411
\bibitem[Voskresensky (2001)]{V00}
Voskresensky, D.N. 2001, in: 'Physics of Neutron Stars Interiors',
Blaschke, D., Glendenning, N.K., Sedrakian, A. (Eds), Lect. Notes
Phys. {578}, 467; [arXiv:astro-ph/0101514]
\bibitem[Voskresensky   \&  Mishustin (1982)]{VM82}
Voskresensky, D.N., Mishustin, I.N. 1982, Sov. J. Nucl. Phys.
{35}, 667
\bibitem[Voskresensky et al. (2000)]{VKZC}
Voskresensky, D.N.,  Khodel, V.A., Zverev, M.V.,   Clark J.W.
2000, Ap.J. {533}, L127; [arXiv:astro-ph/0003172]
\bibitem[Wiringa et al. (1995)]{WSS95}
Wiringa, R.B., Stoks, V.G., Schiavilla, R. 1995, Phys. Rev. C,
{51}, 38
\bibitem[Yakovlev et al. (2004)]{YGKLP03}
Yakovlev, D.G., Gnedin, O.Y., Kaminker, A.D., Levenfish, K.P.,
Potekhin, A.Y. 2004a, Adv. Space  Res., {33}, 523
\bibitem[Yakovlev et al. (2003b)]{YLPGC03}
Yakovlev, D.G., Levenfish, K.P., Potekhin, A.Y., Gnedin, O.Y.,
Chabrier, G.
 2004b, A \& A, {417}, 169
\end{thebibliography}
\end{document}